\documentclass[prb,twocolumn,floats,showpacs,superscriptaddress]{revtex4}

\usepackage{bm}
\usepackage{graphicx}
\usepackage{amssymb}
\usepackage{amsfonts}
\usepackage{amsmath}
\usepackage{epstopdf}
\usepackage{color}

\def \FUW{Institute of Experimental Physics, Faculty of Physics, University of Warsaw, ul. Pasteura 5, 02-093 Warsaw, Poland}
\def \IFPAN{Institute of Physics, Polish Academy of Sciences, al. Lotnik\'{o}w 32/46, 02-668 Warsaw, Poland}
\def \GEMAC{Groupe d'{\'E}tude de la Mati\`{e}re Condens{\'e}e (GEMaC), Universit{\'e} de Versailles
St-Quentin en Yvelines-CNRS, Universit{\'e} Paris-Saclay, Versailles, France}
\def \GIZA{Solid State Physics Department, National Research Center, Giza, Egypt}
\def \UB{Institute of Physics, Kazimierz Wielki University, Powstancow Wielkopolskich 2, 85-064 Bydgoszcz, Poland}

\begin{document}

\title{Fe dopant in ZnO: 2+ $vs$ 3+ valency and ion-carrier \emph{s,p-d} exchange interaction}

\author{J. Papierska} \affiliation{\FUW}
\author{A.~Ciechan} \affiliation{\IFPAN}
\author{P.~Bogus{\l}awski} \affiliation{\IFPAN} \affiliation{\UB}
\author{M. Boshta} \affiliation{\GIZA}
\author{M. M. Gomaa} \affiliation{\GIZA}
\author{E. Chikoidze} \affiliation{\GEMAC}
\author{Y. Dumont} \affiliation{\GEMAC}
\author{A.~Drabi\'nska} \affiliation{\FUW}
\author{H.~Przybyli\'nska} \affiliation{\IFPAN}
\author{A.~Gardias} \affiliation{\FUW}
\author{J. Szczytko} \affiliation{\FUW}
\author{A. Twardowski} \affiliation{\FUW}
\author{M.~Tokarczyk} \affiliation{\FUW}
\author{G.~Kowalski} \affiliation{\FUW}
\author{B. Witkowski} \affiliation{\IFPAN}
\author{K. Sawicki} \affiliation{\FUW}
\author{W. Pacuski} \affiliation{\FUW}
\author{M. Nawrocki} \affiliation{\FUW}
\author{J. Suf\mbox{}fczy\'nski}\email{Jan.Suffczynski@fuw.edu.pl} \affiliation{\FUW}

\date{\today}

\begin{abstract}

Dopants of transition metal ions in II-VI semiconductors exhibit native 2+ valency. Despite this, 3+ or mixed 3+/2+ valency of iron ions in ZnO was reported previously. Several contradictory mechanisms have been put forward for explanation of this fact so far.
Here, we analyze Fe valency in ZnO by complementary theoretical and experimental studies. Our calculations within the generalized gradient approximation (GGA+$U$) indicate that the Fe ion is a relatively shallow donor. Its stable charge state is Fe$^{2+}$ in ideal ZnO, however, the high energy of the (+/0) transition level enhances the compensation of Fe$^{2+}$ to Fe$^{3+}$  by non-intentional acceptors in real samples. Using several experimental methods like electron paramagnetic resonance, magnetometry, conductivity, excitonic magnetic circular dichroism and magneto-photoluminescence we confirm the 3+ valency of the iron ions in polycrystalline (Zn,Fe)O films with the Fe content attaining 0.2\%.We find a predicted increase of $n$-type conductivity upon the Fe doping with the Fe donor ionization energy of $0.25 \pm 0.02$~eV consistent with the results of theoretical considerations. Moreover, our magnetooptical measurements confirm the calculated non-vanishing \emph{s,p-d} exchange interaction between band carriers and localized magnetic moments of the Fe$^{3+}$ ions in the ZnO, being so far an unsettled issue.
\end{abstract}




\maketitle
\section{Introduction}
The magnetically doped zinc oxide holds a great promise for implementations in optoelectronic devices.\cite{Ando:JAP2001, Kittilstved:NatureMat2006, Coey:DiluteMOxides2007, Pacuski:book2010} As a member of a wide bandgap semiconductor family it is characterized by a small lattice constant, a possible large \emph{p-d} hybridization, a small spin-orbit interaction, and a large exciton oscillator strength. These properties make ZnO based dilute magnetic semiconductors DMSs highly attractive for room temperature applications.\cite{Dietl:Science2000, Kittilstved:NatureMat2006, Dietl:NatureMat2010} Unique magnetic and magnetooptical properties have been already demonstrated, e.g., for (Zn,Co)O and (Zn,Mn)O.\cite{Ando:JAP2001, Ando:JPCM2004, Pacuski:PRB2006, Pacuski:PRB2011, Sawicki:PRB2013, Alsaad:PhysBCM2014} Ultra long spin coherence time ($>150~\mu$s) found recently\cite{Tribollet:EPL2008} for Fe$^{3+}$ ions in the ZnO indicates (Zn,Fe)O as a highly promising spintronic system.
Systematic studies of magnetooptical properties of the ZnO doped with iron ions are, however, still missing. In particular, in the work of Ando et al. (Ref.~\onlinecite{Ando:JPCM2004}) Magnetic Circular Dichroism (MCD) spectrum has been shown for the (Zn,Fe)O, but no dependence on the magnetic field nor the iron charge state was determined. Also, there is no literature record related to properties of near-the-band-gap photoluminescence (PL) of the (Zn,Fe)O in magnetic field.

Dopants of transition metal ions in II-VI semiconductors natively exhibit the 2+ valency. In particular, 2+ valency was reported for the iron ions in CdS, ZnS, CdSe, ZnSe, CdTe and ZnTe.\cite{Slack:PR1966, Baranowski:PR1967, Slack:PR1967, Slack:PR1969, Buhmann:PRB1981, Hausenblas:SSC1989, Udo:PRB1992,Smolenski:NatureCommun2016} In the case of ZnO the Fe ions have been, however, observed in parallel in both valence states (2+ and 3+)\cite{Kim:JAP2004, Ahn:JMMM2004, Karmakar:PRB2007, Lin:JofAlloysComp2007, Malguth:pssb2008} or exclusively in 3+ state.\cite{Heitz:PRB1992}

Several contradictory mechanisms have been put forward for explanation of 3+ valency of iron in ZnO, like promotion of Fe$^{2+}$ into the Fe$^{3+}$ due to a compensation induced by Zn vacancies,\cite{Karmakar:PRB2007} photoionization of the Fe$^{2+}$ centers\cite{Heitz:PRB1992} or a direct charge transfer from the ion to the conduction band\cite{Malguth:pssb2008}. In parallel, the long spin coherence time\cite{Tribollet:EPL2008} suggests that the Fe$^{3+}$ ions in ZnO are decoupled from their environment. Thus, there are at least two questions that still need to be answered: (i) What is the valency of Fe ions in ZnO? (ii) Do the Fe$^{3+}$ ions couple to band carriers through \emph{s,p-d} exchange interaction?

In this work, we start with presenting the results of the density functional theory (DFT) study of the electronic and magnetic properties of Fe in ZnO.
The generalized gradient approximation, together with the $+U$ corrections considered as fitting parameters, are employed. Calculations performed with $U$(Fe) larger than 3~eV show that the Fe$^{2+}$  is a relatively shallow donor with the (+/0) transition level in the upper part of the gap, and therefore the stable charge state of Fe in ideal ZnO is the Fe$^{2+}$. The experimental ionization energy, around 0.25~eV, is reproduced with $U$(Fe)~=~4~eV. The high energy of (+/0) level relative to the valence band facilitates compensation of the Fe$^{2+}$ to the Fe$^{3+}$ by, e.g., native acceptors because of the so-called Fermi level effect. When $U$(Fe) lower than 3~eV is assumed, the donor level of the Fe$^{2+}$  is degenerate with the conduction band, while that of the Fe$^{3+}$ forms a gap state. In this case, a correct description of the electronic structure of Fe within GGA is not possible, as discussed in Appendix~\ref{Appendix}. Finally, our calculations point toward non-zero values of \emph{s,p-d} exchange constants in the (Zn,Fe)O.

In the second part of the paper we provide a set of experimental results of magnetometry, conductivity and magnetooptical measurements on the (Zn,Fe)O samples with the Fe content attaining 0.2\%. They confirm the theoretical findings. Magnetometry results point toward 3+ valency of the Fe ions. Conductivity measurements reveal ionization energy of the Fe ions around 0.25~eV. Brillouin-like dependencies found in measurements of MCD and of PL in magnetic field along with a clear Curie-paramagnetic dependence of magnetization on temperature determined from magnetospectroscopy confirm presence of the Fe ions in 3+ charge state and the ion-carrier \emph{s,p-d} exchange interaction in the (Zn,Fe)O.

The paper is organized as follows. Section~\ref{Theory} presents the method and the results of the calculations, which explain the valency of iron ions in ZnO and predict non-zero \emph{s,p-d} exchange integrals for the (Zn,Fe)O. Sec.~\ref{sec:Samples} describes the samples studied along with the results of their structural characterization. Sec.~\ref{sec:Exp:Valency} gathers experimental results obtained with non-optical methods, which testify the presence of iron ions in 3+ valency in the studied samples. Sec.~\ref{sec:Magnetospectroscopy} reports on the results of magnetospectroscopy investigations, which confirm the findings from the Sec.~\ref{Theory} and Sec.~\ref{sec:Exp:Valency}, in particular by providing evidence for the \emph{s,p-d} exchange interaction in the (Zn,Fe)O.

\section{\label{Theory} Theory}

\subsection{\label{teo1}Theoretical background and method of calculations}
Theoretical description of transition metal (TM) impurities in semiconductors represents a demanding test for electronic structure calculations for two reasons. The first one is that the local density approximation (LDA) and GGA to the DFT severely underestimate the band gap, due to which the TM levels can be incorrectly predicted to form resonances degenerate with the conduction band continuum, rather than the experimentally observed states in the band gap. This error can lead to an erroneous charge state of a TM impurity, metallic rather than insulating crystal, spurious magnetic interactions between the dopants, etc.~\cite{t12, t12a} The second problem is related to a localized nature of the 3$d$ wave functions, for which many body effects can play an important role, requiring usage of approaches beyond LDA/GGA. This holds for both orbitals of host semiconductors and $d$(TM) impurity orbitals.
In the case of ZnO, both LDA and GGA give a too small band gap $E_{gap}$ of about 1.0~eV,~\cite{theo2, theo4, t7} and a too high energy of the $d$(Zn)-derived bands relative to the valence band maximum (VBM).~\cite{theo1}

The hybrid functionals~\cite{Uddin, Oba2008, Oba2010, Wrobel} and quasiparticle GW approaches~\cite{schilfgaarde, Schilfgaarde2006, Shishkin, Fuchs, t7, Shih} are more accurate methods to describe the band structure. However, as recently pointed out in Ref.~\onlinecite{t7}, they improve mainly the band gap, while placing the $d$(Zn) band too high. Consequently, Lim {\it et al.} applied an on-site potential for $d$(Zn) states during GW calculations to shift down the $d$(Zn)-derived bands and hence to obtain full agreement with experiment.~\cite{t7}

An alternative pseudoempirical approach consists in using GGA supplemented by the $+U$ corrections,~\cite{t3, t3a, coco} which can be calculated e.g.  by the linear response method~\cite{coco} or can be treated as free parameters fitted to experimental data. The $+U$ term applied to $d$(Zn) orbitals~\cite{theo3, t7, t6, t16} greatly improves the position of $d$ bands, but hardly affects the band gap. The correct gap was obtained by applying an extremely large $U$ term to $s$(Zn) states ($\sim 40$~eV) in addition to $U$ for $d$(Zn) states.~\cite{Paudel, Lany2008} Another method consisted in using $+U$ for $d$(Zn) supplemented by empirical non-local external potentials on $s$ and $p$ states of both Zn and O atoms.~\cite{t12a}
To obtain the correct $E_{gap}$ within GGA+$U$ one should note that the upper valence band is mainly derived from the $p$(O) orbitals. Consequently, the $U$ term for the $p$(O) orbitals, in addition to $U$(Zn), should be included.~\cite{t5, Calzolari, marco} This approach is employed in this work, as described below.

The calculations are performed within DFT theory with the GGA for the exchange-correlation potential.~\cite{t1, t2}
The $+U$ corrections are included according to Refs~\onlinecite{t3, t3a, coco}.
We use the pseudopotential method implemented in the {\sc Quantum ESPRESSO} code,~\cite{t4} with the valence atomic configuration $3d^{10}4s^2$ for Zn, $2s^2p^4$ for O and $3s^2p^6 4s^2p^0 3d^6$ for Fe, respectively.
The plane-waves kinetic energy cutoffs of 30~Ry for wave functions and 180~Ry for charge density are employed.
The electronic structure of ZnO in the wurtzite phase is examined with the $8\times 8\times 8$ $k$-point grid. Analysis of a single Fe impurity in ZnO is performed using the $3\times 3\times 2$ supercell with 72 atoms (2.8 at \% of Fe) and the $3\times 3\times 4$ supercell with 144 atoms (1.4 at \% of Fe). For the density of states (DOS) calculations, the $k$-space summations are performed with a $3\times 3\times 3$ $k$-point grid.
Ionic positions are optimized until the forces acting on ions are smaller than 0.02~eV/\AA. Methfessel-Paxton~\cite{theo_MP} smearing method with the smearing width of 0.136~eV or lower is used for partial occupancies. Calculations with fixed occupation matrices are performed at the $\Gamma$ point only using the smaller $3\times 3\times 2$ supercell.

The values of the $U$ terms for 3$d$(Zn) and 2$p$(O) orbitals are fitted to reproduce the band structure of ZnO. We find that $U$(Zn)=12.5~eV and $U$(O)=6.25~eV reproduce both the experimental $E_{gap}$ of 3.3~eV,~\cite{t6, Izaki:APL1996, Srikant:JAP1998} and the energy of the $d$(Zn) band, centered about 8~eV below the VBM.~\cite{t7}
They also lead to the proper width of 6~eV of the upper valence band of mostly $p$(O) character, in accordance with the experiment.~\cite{t7}
Our $U$ parameters, although relatively high, are similar to the values reported in other works. In Refs~\onlinecite{t5} and~\onlinecite{Calzolari}, $U$(Zn)$=10-12$~eV and $U$(O)$=6-7$~eV were fitted to the experimental band structure. Importantly, in Ref.~\onlinecite{marco}, parameters $U$(Zn)=12.8~eV and $U$(O)=5.29~eV were calculated by using pseudohybrid Hubbard density functional method. This is consistent with the fitted values, and thus provides a complementary justification of the used $+U$ parameters.

The relaxed crystal structure agrees well with experiment:
the lattice parameters $a= 3.23$~\AA\ and $c = 5.19$~\AA\ as well as internal parameter $u = 0.38$ are underestimated by less than 1~\% in comparison with experimental values $a =3.25$~\AA, $c = 5.20$~\AA, and $u = 0.38$.~\cite{t8}
The value of $U$(Fe) is considered as a free parameter varying from 0 to 6~eV. Finally, the increase of the temperature from 0 to 300~K causes a change of the measured band gap by about 0.1~eV. This change is neglected in the calculations, since it does not affect the conclusions.

\subsection{\label{teo2} Energy levels of Fe impurity in ZnO}
In ideal II-VI semiconductors, i.e., those without additional dopants and defects, a single Fe impurity is expected to occur in the Fe$^{2+}$ ($q=0$) charge state, with 6 electrons on the $d$(Fe)-induced levels and total spin $S=2$. However, as it follows from the present results, due to proximity of Fe levels to the ZnO conduction band, the description of electronic configuration of Fe in ZnO is not straightforward. More specifically, for small values of $U$(Fe), $0 \leq U$(Fe)$\leq 2.5$~eV, the calculated level of Fe$^{2+}$  is degenerate with the continuum of the conduction band, and there are fundamental problems with the correct description of this case within LDA/GGA. This issue is discussed in detail in Appendix A. On the other hand, those problems are absent when higher values of $U$(Fe) are assumed. In this case, in the ideal ZnO, Fe is a standard impurity in the 2+ charge state. Due to the proximity of the donor level to the conduction band we predict, however, an easy ionization of Fe$^{2+}$  to the Fe$^{3+}$, either by thermal excitation of an electron to the conduction band, or by electron transfer to acceptors present in the host. This means that in the realistic case of imperfect samples both Fe$^{2+}$  and Fe$^{3+}$ should co-exist.

\begin{figure}[t]
\begin{center}
\includegraphics[width=1\linewidth]{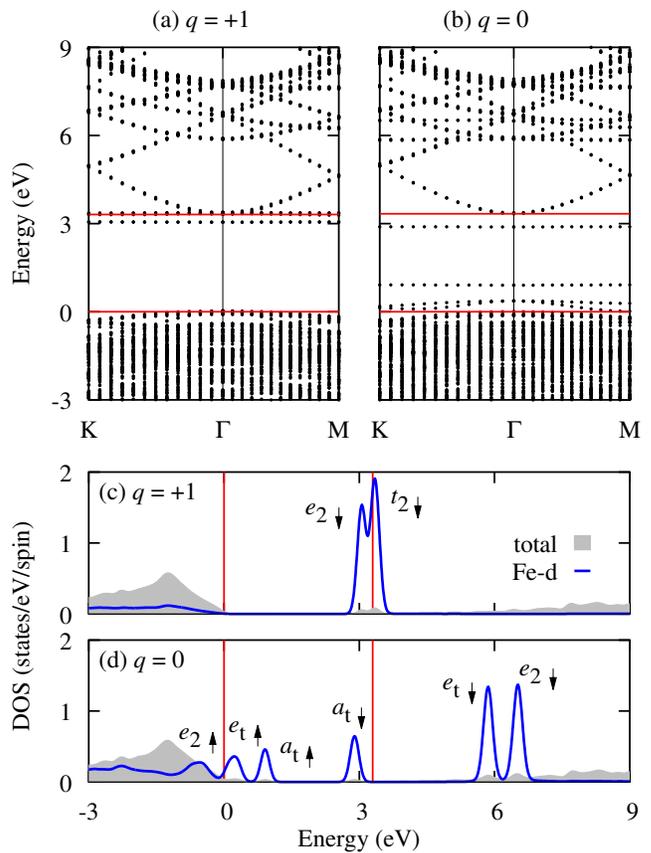}
\caption{\label{figT1}
(Color on-line)
Energy bands and DOS of ZnO with Fe in: (a) and (c) $q=+1$ charge state, Fe$^{3+}$, (b) and (d) $q=0$ charge state, Fe$^{2+}$. Grey area and blue lines in DOS indicate the total DOS and the DOS projected on $d$(Fe) orbitals, respectively. Red lines denote the band gap of ZnO. Zero energy is set at the VBM. The calculation is performed with the 144-atom supercell for $U$(Fe) = 4~eV.}
\end{center}
\end{figure}
The band structure and DOS of ZnO calculated assuming $U$(Fe)$=4$~eV is shown for the Fe$^{3+}$ and Fe$^{2+}$ case in Fig.~\ref{figT1}(a) and Fig.~\ref{figT1}(b), respectively.
The impurity $d$(Fe) gap states constitute nearly dispersionless bands, while their energies strongly depend on the Fe charge state. To make the discussion transparent we begin the analysis by Fe$^{3+}$, and then move to Fe$^{2+}$.

The Fe ion in the $q=+1$ charge state, Fe$^{3+}$, assumes the electronic configuration $d^5$ characterized by 5 electrons on the $d$(Fe) levels and total spin $S=5/2$. In this case, Fe introduces two spin-up states below the VBM: the $e_{2\uparrow}$ doublet and the $t_{2\uparrow}$ triplet, which form broad resonances within the valence band, see Fig.~\ref{figT1}(c). The exchange spin-up$-$spin-down splitting is strong, $\sim 3$~eV, since the $e_{2\downarrow}$ doublet is at 3.0~eV, while the $t_{2\downarrow}$ triplet is practically degenerate with CBM. Both spin down levels are empty.
Addition of one electron to the ZnO:Fe$^{3+}$ system results in the $q=0$ charge state  Fe$^{2+}$, with 6 electrons on the $d$(Fe)-induced levels. In this case, the energies of Fe-induced levels rise due to the increased intracenter Coulomb repulsion, the $e_{2\uparrow}$ doublet is still degenerate with the valence band, while the $t_{2\uparrow}$ triplet is at about 0.8~eV above VBM, and is split into a $e_{t\uparrow}$ doublet and a $a_{t\uparrow}$ singlet. More importantly, the $a_{t\downarrow}$ singlet occupied with one electron is 0.25 eV below CBM.

\begin{figure}[t!]
\begin{center}
\includegraphics[width=8.3cm]{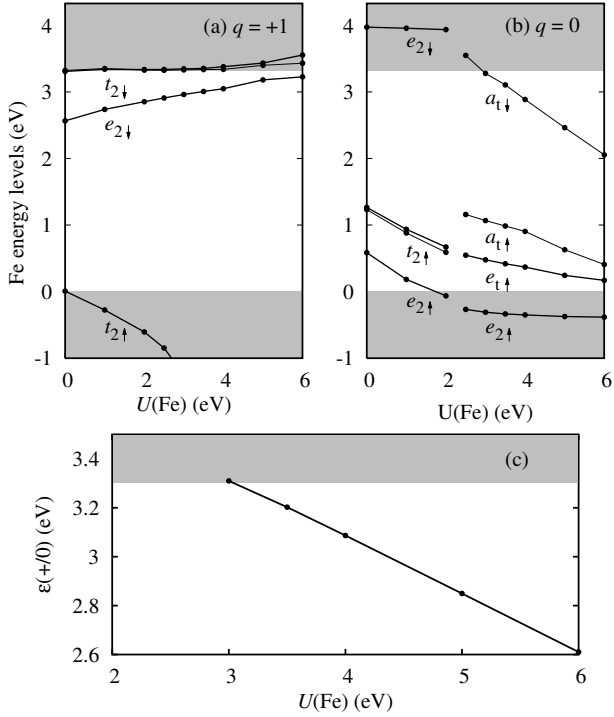}
\caption{\label{figT2}
(Color on-line) Energy levels of (a) Fe$^{3+}$, (b) "Fe$^{2+}$" (for $U$(Fe)$\leq 2.5$~eV, see Appendix A) and Fe$^{2+}$ (for $U$(Fe)$ > 3$~eV), and (c) the transition level $\varepsilon (+/0)$ calculated as a function of $U$(Fe). The 72 atom supercell is used in calculations.}
\end{center}
\end{figure}
The value of $U$(Fe) is treated here as a parameter to be fitted to experiment. The dependence of the Fe states on the $U$(Fe) is presented in Fig.~\ref{figT2}. As it is discussed in Refs~\onlinecite{VZB, ZB} for transition metal impurities in GaN, the $U$-induced shift of an impurity level depends on its occupation, and is negative (positive) for the occupied (empty) state. This feature is clearly seen in Fig.~\ref{figT2}(a) for both the occupied $e_{2\uparrow}$ and $t_{2\uparrow}$, and the empty $e_{2\downarrow}$ and $t_{2\downarrow}$ levels of Fe$^{3+}$. The calculated dependencies of $e_{2\uparrow}$ and $t_{2\uparrow}$ on $U$(Fe) are non-linear because of their increasing hybridization with valence states.

The case of Fe$^{2+}$, Fig.~\ref{figT2}(b), is more complex.
For the $U$(Fe) up to 2.5~eV, the calculated energies of the spin down levels are degenerate with the conduction band, but this  "pseudo-resonant" character of Fe is largely an artefact of GGA, see Appendix~\ref{Appendix}. The $t_{2\downarrow}$ triplet is about 1 eV above $e_{2\downarrow}$, as for Fe$^{3+}$. In the interval $2 <U$(Fe)$<2.5$~eV, an abrupt change in the order the Fe-induced levels occurs. In particular, the $t_{2\downarrow}$ triplet splits, the $a_{t\downarrow}$ singlet derived from $t_{2\downarrow}$ is lower in energy than the $e_{2\downarrow}$ doublet, becoming a gap state for the $U$(Fe) higher than 3~eV (see Ref.~\onlinecite{ref_occupancy}). Since $a_{t\downarrow}$ is occupied with one electron, its energy decreases with the $U$(Fe).
This splitting of $t_{2\downarrow}$ changes the atomic configuration by increasing the
difference between the length of the Fe-O bond along the $c$-axis and those of the three remaining planar bonds, see Ref.~\onlinecite{bonds}. This, in turn, induces the splitting of the $t_{2\uparrow}$ gap state for $U$(Fe)$>2.5$~eV. Consequently, for the $U$(Fe)$>3$~eV, Fe is a donor with an occupied singlet in the band gap and spin 2. This allows for calculations of the transition level and ionization energy of Fe.

The change of the charge state of Fe is determined by the $\varepsilon (+/0)$ transition level between $q=+1$ and $q=0$ charge states. It is defined as the Fermi energy relative to VBM at which formation energies of the 0 and +1 charge states are equal:
\begin{equation}
\varepsilon(+/0)={E(q=0)-E(q=+1)}-e_{VBM},
\end{equation}
where $E(q)$ is the total system energy. The calculations are performed along the scheme proposed in Ref.~\onlinecite{Lany2008}. The energy of VBM, $e_{VBM}$, is determined from the total energy difference between the pure ZnO crystal with and without a hole in the VBM, i.e.
\begin{equation}
e_{VBM}=E(ZnO,q=0)-E(ZnO,q=+1).
\end{equation}

\begin{figure}[t!]
\begin{center}
\includegraphics[width=8cm]{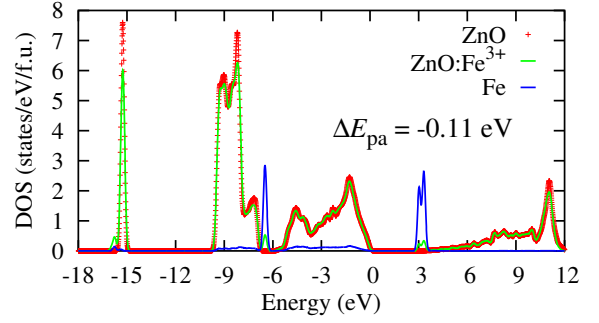}
\caption{\label{figT3}
(Color on-line)
DOS of pure ZnO and ZnO with Fe in the $q=+1$ charge state, Fe$^{3+}$, after potential alignment $\Delta E_{pa} = -0.11$~eV. To improve visibility, a contribution of the Fe states, shown in blue, is multiplied by 10.
$U$(Fe) = 4~eV is used in calculations.
}
\end{center}
\end{figure}
In general, the supercell results must be corrected for finite size effects by taking into account the band filling correction, and the image charge corrections together with potential alignment for charged defects.~\cite{Lany2008,Lany2009,VdW} In the latter case,
a compensating homogeneous background charge is assumed. This guarantees charge neutrality, i.e., a correct value of the ${\bf G}=0$ component of electrostatic potential, but it also requires an alignment of the average electrostatic potential in the supercell with and without the defect. The potential alignment energy $\Delta E_{pa}$  is estimated by comparing the DOS of pure ZnO and of ZnO:Fe$^{3+}$. Far from the defect levels, the energy shift for charged system is $\Delta E_{pa}=-0.11$~eV (Fig.~\ref{figT3}).

Image charge correction $\Delta E_{MP}$ stems from the electrostatic coupling between charged defects in different supercells. The simplified version of the one proposed in Ref.~\onlinecite{Lany2009} is employed,
\begin{equation}
\Delta E_{MP} = [1 + c_{sh} (1 - \varepsilon^{-1} )] \frac{q^2\alpha_M}{2\varepsilon
V^{1/3}},
\end{equation}
\noindent
where $\alpha_M$ is the Madelung constant, $\varepsilon=10.3$ is the low-frequency dielectric constant, $V$ is the supercell volume, and $c_{sh}=-0.365$ is the shape factor for a $3\times 3\times 2$ HCP supercell. This gives $\Delta E_{MP}=0.16$~eV. Thus, the potential alignment and the image charge term cancel to a good approximation. Regarding the convergence with respect to the supercell size we note that supercells with about 100 atoms give well convergent results even for highly charged defects, see Ref.~\onlinecite{Lany2009}. Our results confirm this, since the values obtained with 72 and 144 atom supercells agree to better than 0.1~eV. Finally, in the present case the band filling correction vanish, since the bands derived from the $d$(Fe) are fully occupied for both Fe$^{2+}$ and Fe$^{3+}$.

The calculated $\varepsilon (+/0)$ transition level is shown in Fig.~\ref{figT2}(c) as a function of $U$(Fe). In agreement with Fig.~\ref{figT2}(b), $\varepsilon (+/0)$ coincides with the CBM for $U$(Fe)=3~eV, and decreases in energy with the increasing $U$(Fe).

A quantity often considered in the context of experiment is the ionization $E_{ion}$ energy. (Here, the $E_{ion}$ is experimentally determined from the temperature dependence of conductivity in  Sec.~\ref{sec:Transport}.) It can be calculated as the total energy difference between the final and the initial states of the system, which are the Fe$^{3+}$ with one electron in the conduction band and Fe$^{2+}$, respectively. As it is pointed out in Refs~\onlinecite{Lany2008, VdW}, the energy of the $\varepsilon (+/0)$ level counted from the VBM and the thermal ionization energy $E_{ion}$ should add up to the $E_{gap}$. In practice, when the two quantities are determined in calculations small differences between them occur due to the used approximations.
For instance, when calculating the total energy of the $q=+1$ state, the fictitious neutralizing background is assumed. (However, in actual calculations the background is not introduced. Instead, the ${\bf G}=0$ component of the electrostatic potential is assumed to vanish.~\cite{Lany2009})
On the other hand, in the calculations of the $E_{ion}$, both in the initial and the final state the supercell are electrically neutral, and introduction of the background (or the adjustment of the $\mathbf{G}=0$ component of the total electrostatic potential) is not necessary. Another source of discrepancies is the limitation of the Brillouin zone summations to the $\Gamma$-point, used in the calculations of excited states with fixed non-equilibrium occupations. In our case, both approaches agree within 0.1~eV.

By comparing theory with experiment we find that our experimental value $E_{ion}=0.25\pm.02$~eV is reproduced when $U$(Fe)=4~eV, Fig.~\ref{figT2}(c).~\cite{note4}
This is close to 4.3~eV for FeO, and larger than $U$(Fe)$=2.2\pm 0.2$~eV found for bulk Fe.~\cite{coco}
It is worth to note that despite the almost vanishing $U$ for Mn and Fe found in the corresponding case of GaN,\cite{VZB, ZB} a recent study~\cite{CB_unpublished_Mn} indicates that also for Mn in ZnO, the $U$(Mn) term is about 1.5~eV, i.e., larger than in GaN.

Previous theoretical investigations of Fe in ZnO provided conflicting results regarding the stability of the Fe$^{2+}$ charge state.~\cite{sandra, park, spaldin, t11, t12, t14} The uncertainty is largely due to the band gap problem, mentioned in the Subsection~\ref{teo1}. The LDA/GGA calculations typically situated the $e_{2\downarrow}$ of Fe$^{2+}$ above the CBM, but this level ordering can result from the too low band gap. On the other hand, the gap-corrected approaches predicted the $e_{2\downarrow}$ state below the CBM.
In the former case, the instability of Fe$^{2+}$ is expected as discussed above, but this possibility was not explicitly addressed. More specifically, in Ref.~\onlinecite{sandra} the LDA was used. The $e_{2\downarrow}$ state of Fe was found to pin the Fermi level, and its energy was degenerate with the CBM; this is close to the present results for small $U$(Fe), see Appendix~\ref{Appendix}. Such a situation was also predicted by the LDA calculations of Ref.~\onlinecite{park}. LDA was also used in Ref.~\onlinecite{spaldin} with the $+U$ term applied only to $d$(Fe) orbitals, which gave a strongly underestimated band gap, and the energy of $e_{2\downarrow}$ above the CBM. Previous theoretical approaches to Fe in ZnO with the corrected gap included the LDA with the self-interaction corrections and $U$(Fe)$=5.5$~eV;~\cite{t11} this provided an Fe-induced spin-down band at 1.8~eV above the VBM. Similarly, in Ref.~\onlinecite{t12} the band structure of pure ZnO was fitted to experiment, and the corrections for $d$(Fe), $U=3.5$~eV and the exchange constant
$J=1$~eV (thus effectively $U$(Fe)=2.5~eV) were applied. The $(+/0)$ transition level was found at 1.98~eV, i.e., about 1.4~eV below the CBM. In Ref.~\onlinecite{t14}, GGA$+U$ with $U$(Fe)$=2.2$~eV and $U$(Zn)$=4.7$~eV was employed together with an empirical band gap correction. The $(+/0)$ transition level was predicted at 2.9~eV above the VBM. This is close to our results with $U$(Fe)$=4-5$~eV.

\subsection{\label{teo3}\emph{s,p-d} coupling}
\begin{table}
\caption{\label{table1}
The Fe magnetic moment $M$~($\mu_B$), conduction $\Delta E_c$~(eV) and valence $\Delta E_v$~(eV) band-edge spin splittings, and exchange constants $N_0\alpha$~(eV), $N_0\beta$~(eV) for $q=0$ and $q=+1$ charge state. Results are obtained for the 72-atom supercell, $x=0.028$. For $q$=+1 and $U$(Fe)=0, values of the $\Delta E_v$ and the $N_0\beta$ can not be precisely determined, as indicated by "$\ast$" (see Sec.~\ref{teo3} for details).
}
\begin{ruledtabular}
\begin{tabular}{l l c c c c c }
       &$U$(Fe)  & $M$   &$\Delta E_c$ &$\Delta E_v$ &$N_0\alpha$ &$N_0\beta$\\ \hline
$q=0$  &0        & 4.29  &0.024        & 0.033        &0.41      &0.55 \\
       &4~eV     &4.00   &0.032        &0.112         &0.57      &2.01 \\ \hline
$q=+1$ &0        & 4.89  &0.030        & $\ast$       &0.44      &$\ast$ \\
       &4~eV     & 5.00  &0.027        &-0.029        &0.38      &-0.42 \\
\end{tabular}
\end{ruledtabular}
\end{table}

The calculated spin splitting of the conduction band $\Delta E_c = E_{c\downarrow} - E_{c\uparrow}$ and the valence band $\Delta E_v = E_{v\downarrow} - E_{v\uparrow}$, produced by the coupling of Fe with the host ZnO states, can be used to estimate the \emph{s,p-d} coupling. Within the mean-field approximation, the exchange constants are expressed by~\cite{Sanvito}
\begin{equation}
N_0\alpha=\Delta E_c/(x \langle S \rangle), \quad
N_0\beta=\Delta E_v/(x \langle S \rangle),
\end{equation}
where $x$ is the concentration of the Fe ions and $\langle S \rangle$ is one half of the computed magnetization.

The calculated values are given in Table~\ref{table1}. As it follows from the Table, the constant $N_0\alpha$ is around $0.4$~eV for both, 2+ and 3+, Fe charge states. The obtained value is comparable to the typical one, 0.2~eV, found in II-VI compounds.~\cite{kacman} Moreover, the $N_0\alpha$ depends on the $U$(Fe) only weakly.

In contrast, the constant $N_0\beta$ is strongly dependent on both the Fe charge state and the $U$(Fe). For $q=0$ and $U$(Fe)$\leq 2$~eV, the $N_0\beta$ is obtained for "Fe$^{2+}$" charge state, and thus it should be considered as an estimate.
The obtained $N_0\beta$, $0.5-1$~eV, is  smaller than typical iron-hole exchange integrals for II-VI compounds,~\cite{Scalbert:SSC1990,Twardowski:PRB1990, Testelin:SSC1991,Testelin:SSC2000} but of the same order as effective exchange integrals reported for wide gap DMSs~\cite{Pacuski:PRB2006, Przezdziecka:SSC2006, Pacuski:PRB2007, Pacuski:PRL2008, Dietl:PRL2008, Pacuski:PRB2011, Suffczynski:PRB2011}
With the increasing $U$(Fe), $t_{2\uparrow}$ approaches the valence band (see Fig.~\ref{figT2}), which increases the spin splitting of the VBM, and for the $U$(Fe)$=4$~eV the $N_0\beta$ reaches 2.0~eV .
On the other hand, for $q=+1$, \emph{i. e.}, the Fe$^{3+}$ case, evaluation of the $N_0\beta$ is obscured by energetic proximity of the VBM and the $t_{2\uparrow}$ level. Indeed, as it follows from DOS shown in Fig.~\ref{figT1}(c), for $q=+1$ the $t_{2\uparrow}$ state is degenerate with the top of the valence band, and therefore its coupling with the host valence states is very pronounced. Also a direct inspection of the wave functions of the states close to the VBM reveals the strong hybridization, which makes it impossible to clearly distinguish between band states and Fe-induced states. For the $U$(Fe)=4~eV, the $t_{2\uparrow}$ is well below the VBM, and we obtain the $N_0\beta = -0.42$~eV.

Regarding the sign of the interaction, we find that conduction electrons are ferromagnetically coupled with the Fe impurities, as expected for the direct exchange coupling.~\cite{kacman} The Fe-hole coupling has the kinetic exchange character,~\cite{kacman} driven by the hybridization of the Fe and host states.
The sign of the coupling constant is determined by the energy of $t_{2\uparrow}$ relative to the VBM. For $q=0$, $t_{2\uparrow}$ is above the VBM, and the coupling of holes with the Fe is ferromagnetic. For $q=+1$ the coupling changes the sign. In this case, $N_0\beta$ is formally evaluated from the spin splitting of the highest valence states. Such an approach corresponds to the interpretation of luminescence experiments, in which one determines the spin splitting of the highest valence states involved in the exciton recombination regardless of their actual orbital composition.

The coupling between the Fe and O first nearest neighbors induces magnetic moment on O ions. In the case of Fe$^{3+}$ observed in EPR, the calculated magnetic moments of O ions are $\sim 0.19\ \mu_B$, and that of the Fe is $\sim 4.17\ \mu_B$. The corresponding total magnetic moment is $5.0\ \mu_B$. In the case of Fe$^{2+}$, the spin of the sixth $d$(Fe) electron is antiparallel to the remaining ones, thus both the total magnetic moment and the contributions from Fe and O neighbors are lower than for the Fe$^{3+}$ case. The results hardly depend on the $U$(Fe) value.

\begin{figure}[t!]
\begin{center}
\includegraphics[width=8.3cm]{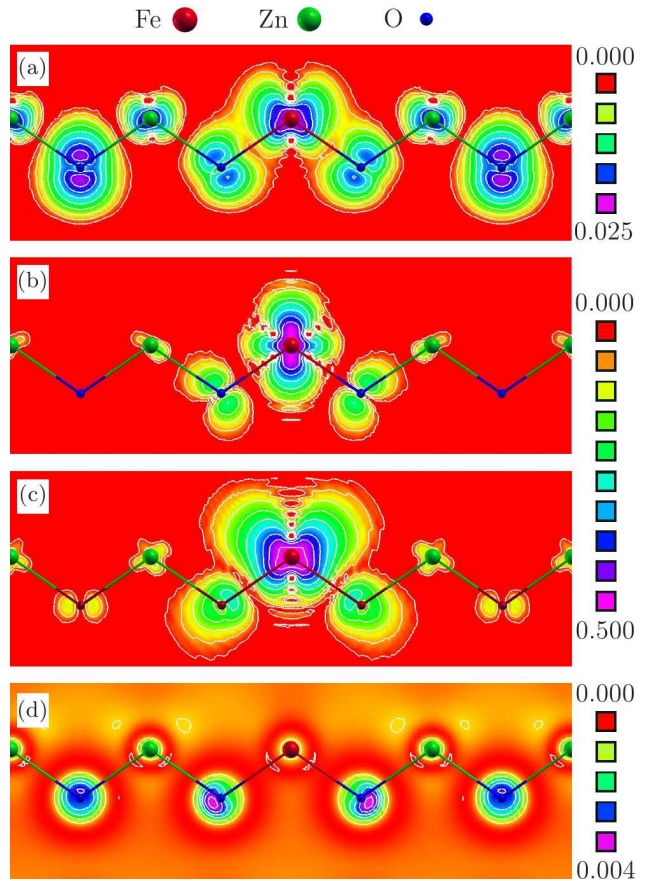}
\caption{\label{figT4}
(Color on-line)
Contour plots of the wave functions squared of (a) the top of the valence band, (b) $e_{2\uparrow}$, (c) $t_{2\uparrow}$, and (d) the bottom of the conduction band. Blue, green, and red dots denote O, Zn, and Fe atoms, respectively.
}
\end{center}
\end{figure}
The wave functions at the $\Gamma$ point of the VBM, $e_{2\uparrow}$, $t_{2\uparrow}$, and the CBM are shown in Fig.~\ref{figT4} for $q=0$ and the $U$(Fe)$=0$. As expected, the top of the valence band is composed of $p$(O) and of $d$(Fe), while the contribution of $d$(Zn) is less pronounced. The two gap states are dominated by $d$(Fe), but the contribution of the host states, mostly the O nearest neighbors, is clearly visible. The hybridization is more pronounced in the case of $t_{2\uparrow}$. (The structure of both $e_{2\uparrow}$ and $t_{2\uparrow}$ is the same as for Mn in GaN,~\cite{Mahadevan, Wolos:PRB2004} while the contribution of the Fe to the VBM in ZnO is lower than that of the Mn in GaN.) The wave functions of the two Fe-induced $e_{2\downarrow}$ and $t_{2\downarrow}$ states, degenerate with the conduction band continuum, are very similar to those of the spin up $e_{2\uparrow}$ and $t_{2\uparrow}$ (see Fig.~\ref{figT4}) and thus are not shown. Finally, the conduction band edge, formed by $s$(O) states and a small contribution of $s$(Zn/Fe) states, is practically not perturbed by the $d$(Fe). The coupling with the conduction states occurs $via$ the direct exchange mechanism,~\cite{kacman} and does not necessitate hybridization. The calculations performed for the $U$(Fe)= 4~eV provide comparable results.

Summing up the theory part of the paper, there are fundamental difficulties with a correct GGA description of the $q=0$ charge state.
They stem from the fact that the $e_{2\downarrow}$, which should be occupied by one electron in the Fe$^{2+}$ case, is above the empty conduction band. In consequence, self-ionization should occur. However, when the electron occupies the CBM, the $e_{2\downarrow}$ level is below CBM, which also does not correspond to the ground state configuration of the system. The paradoxal situation, analogous to that found for the isolated Fe atom~\cite{Janak, Koch}, is analyzed in Appendix A, where it is indicated that a correct description of ZnO:Fe is not possible within GGA.

This problem is absent for both the $U$(Fe)$>2$~eV, and for the $q=+1$ charge state.
In particular, the large $U>3$~eV correction induces a decrease of the $a_{t\downarrow}$ state of Fe$^{2+}$ below the CBM, which corresponds to a standard configuration of a relatively shallow donor with a well-defined transition level in the upper part of the band gap. This facilitates compensation by native acceptors, such as zinc vacancies~\cite{Janotti_PRB76, theo_Lany, Oba2008} thanks to the self-compensation effect ({\it i.e.}, to the dependence of formation energy on the Fermi energy).~\cite{Look}
The Fe$^{3+}$-induced levels are about 1.2~eV lower than those of the Fe$^{2+}$. This large difference in the level energies of $q=0$ and $q=+1$ stems from the strong intracenter Coulomb repulsion between the $d$(Fe) electrons caused by the localization of their wave functions. Interestingly, in the case of Mn in ZnO the intracenter Coulomb repulsion is of comparable strength, and it leads to metastability of the photoexcited Mn$^{3+}$.~\cite{CB_unpublished_Mn}

The exchange coupling constant $N_0\alpha$, about 0.4~eV, does not depend on the Fe charge state. $N_0\beta$ is non-vanishing as well, but it depends on the charge state of the Fe: it amounts to about 2.0~eV for the Fe$^{2+}$, while for the Fe$^{3+}$ it is predicted to be lower and of opposite sign, $-0.4$~eV, since the $t_{2\uparrow}$ level is degenerate with the VBM.


\section{\label{sec:Samples} Samples}
Studied ZnO layers doped with the Fe ions are produced by a spray pyrolysis method on quartz or glass substrates,\cite{Chikoidze:JAP2013} with a respective thicknesses of $600 \pm 50$~nm and $150 \pm 30$~nm, as determined by Scanning Electron Microscope measurements (not shown). The layers exhibit a polycrystalline structure with $100 \pm 20$~nm grain size. As determined by magnetometry measurements (see Sec.~\ref{sec:SQUID}), the atomic concentration of the iron ions is $x = 0.2 \pm 0.05$ \%, which corresponds to the iron ion density of $(5 \pm 1) \cdot 10^{19}$ at/cm$^3$. A layer of pure ZnO is also grown in the same conditions as the Fe doped ones.

\begin{figure}
\includegraphics[width=\linewidth]{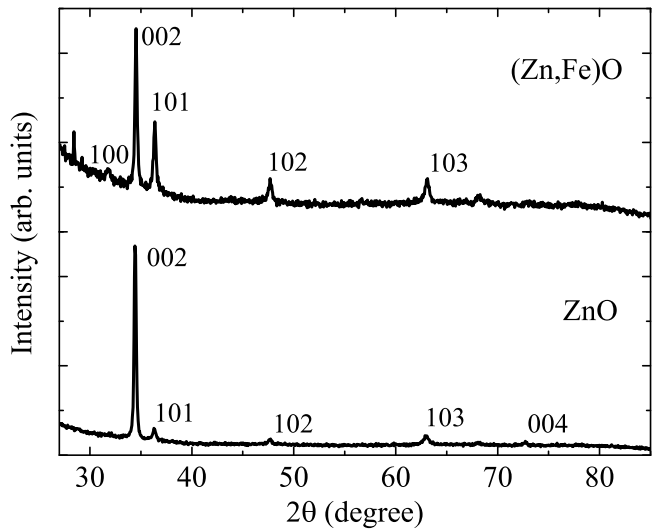}
\caption{X-ray diffraction spectra of (Zn,Fe)O and ZnO layers deposited on quartz. The spectra are vertically shifted for clarity.} 
\label{fig:XRD}
\end{figure}

X-ray diffraction (XRD) characterization shows that the samples exhibit the crystalline structure of the hexagonal ZnO (see Fig.~\ref{fig:XRD}). The Bragg peak (002) is dominant, suggesting that the (002) plane growth rate is the fastest one. The plane (002) is arranged parallel to the substrate, i. e., the preferential c-axis orientation is perpendicular to the film plane for all the samples.


\section{\label{sec:Exp:Valency} Experiment: 3+ valency of iron dopant in {ZnO}}
In order to determine the valence state of iron ions incorporated into the samples, we investigate magnetic and conductivity behavior of the studied layers.

\subsection{\label{sec:EPR} Electron Paramagnetic Resonance}

Electron Paramagnetic Resonance (EPR) measurements are performed at the room temperature using a Bruker ELEXSYS E580 CW spectrometer operating at a microwave frequency around 9.4 GHz (X band) with a TE$_{102}$ resonance cavity. To find the best measurement conditions, the spectra are measured as a function of the microwave power in the range between 0.047~mW and 150~mW, at temperature from 2~K up to the room temperature, and modulation amplitude of the magnetic field from 0.1~mT up to 0.5~mT. The signal from the quartz or glass substrate alone, \emph{i.e.}, without any deposited (Zn,Fe)O layer is also measured and then subtracted as the background from the total signal measured on the studied samples. No powdering of the samples is done.

\begin{figure}
\includegraphics[width=1\linewidth]{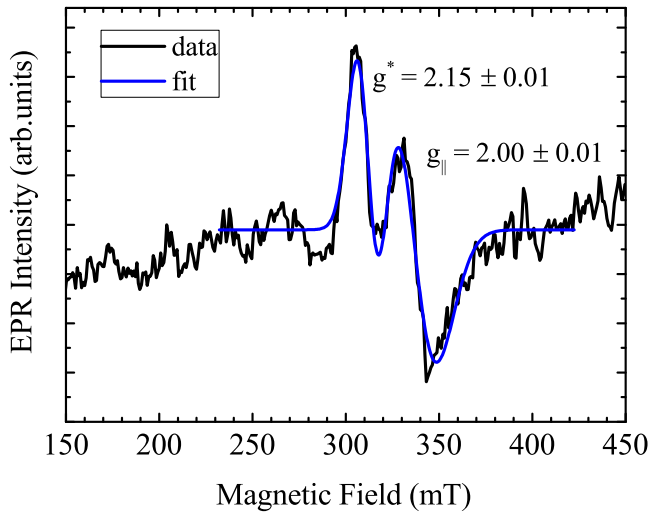}
\caption{
(Color on-line) Electron Paramagnetic Resonance spectrum of (Zn,Fe)O ($x$ = 0.2 \%) sample registered at T = 300 K, acquired with the microwave power 1.5~mW, modulation amplitude 0.1~mT, and microwave frequency 9.393668 GHz. Due to a very weak EPR signal intensity, in order to improve signal-to-noise ratio, the spectrum has been averaged over ten acquisitions. Fit with a derivative of a sum of two Gaussian curves (blue line) yields g$_{||}$ and g* values, as indicated.} \label{fig:EPR}
\end{figure}

Two relatively sharp resonance lines are observed at magnetic fields around 300~mT, superimposed on a broad band spectrum, as shown in Fig.~\ref{fig:EPR}. The fit with a derivative of a sum of two Gaussian curves yields g-factor equal g$_{||}$ = 2.00 $\pm$ 0.01 and g* = 2.15 $\pm$ 0.01, for a resonance respectively at a higher and lower magnetic field value (see Fig.~\ref{fig:EPR}). The g$_{||}$ value agrees very well with the one reported in a previous study for the $-1/2 \leftrightarrow 1/2$ fine structure transition of isolated, substitutional Fe$^{3+}$ ion (S=5/2) in ZnO powders (2.0062).\cite{Acikgoz:JPCM2014} We therefore attribute it to the Fe$^{3+}$ ions in polycrystallites with the c-axis oriented along the normal to the sample plane. We attribute the second resonance and corresponding g-factor g* to $1/2 \leftrightarrow -1/2$ transition of the Fe$^{3+}$ ions in polycrystallites, which c-axes deviate from the normal to the sample plane. A presence of such polycrystallites in the studied samples has been revealed by maxima in XRD spectrum different then 002 (see Fig.~\ref{fig:SQUID}).
Since the linewidth of the weaker $\pm 5/2 \leftrightarrow \pm 3/2$ and $\pm 3/2 \leftrightarrow \pm 1/2$ fine structure transitions is much more sensitive to variations of the polycrystallite size and position of the Fe ion (on the surface of/inside the polycrystallite), the corresponding resonances are smeared out and contribute only to the broad band spectrum. The smearing could be also due to a weak signal from the Fe ions (resulting from their low concentration) and a relatively strong background signal of the substrate, which decreased a signal-to-noise ratio. The impact of spin-spin interaction is most likely negligible here, as the concentration of the iron ions is relatively low.

The agreement of the measured EPR spectrum with that expected for the Fe$^{3+}$ ion in ZnO\cite{Acikgoz:JPCM2014} confirms the presence of the iron ions with the 3+ valency in the studied samples. We note that 2+ ions are not detected in the EPR measurements, and in consequence their presence can not be verified using this method. However, as will be shown below, a good description of the dependence of magnetization and of the MCD on magnetic field is obtained using a Brillouin function as for the Fe$^{3+}$ ions.

\subsection{\label{sec:SQUID} Magnetization}
Magnetization measurements are performed with the use of a SQUID-type magnetometer (liquid helium cooled MPMS XL device manufactured by Quantum Design, providing the sensitivity of $10^{-8}$ emu) in the temperature range 2-300 K and magnetic fields up to 7~T.
The measured magnetic moment of each sample is a sum of paramagnetic moment of the Fe$^{3+}$ ions, a diamagnetic contribution from the ZnO layer and the substrate, and a possible contribution of unintentional impurities. Therefore the measured magnetization can be expressed in the form:
\begin{equation}
M_{exp}(B, T) = M_{(Zn,Fe)O}(B, T) + \chi_{dia} B + C,
\label{Eq1:magnetization}
\end{equation}
where $M_{(Zn,Fe)O}(B, T)$ is a total magnetic moment of the Fe$^{3+}$ ions in ZnO matrix, $\chi_{dia}$ is the sum of diamagnetic susceptibility of ZnO layer and of the substrate (assumed to be temperature independent in the studied temperature range) and $C$ represents a contribution from possible precipitates of secondary phases.

We note that in our case diamagnetic contribution dominates the others since the mass of the magnetic layer is only a tiny fraction of the total mass of the sample. In such a case a precise value of $\chi_{dia}$ of the sample is crucial. A careful analysis of the data reveals the absence of ferromagnetic secondary phases (e.g., Fe-rich aggregates) in all of the studied samples. This is in contrast to what was reported previously for the (Zn,Fe)O\cite{Soumahoro:ThinSolidFilms2009} and other Fe-doped wide band gap semiconductors, e.g., for (Ga,Fe)N.\cite{NavarroQuezada:PRB2010, Dietl:RevModPhys2015} Consequently, the term $C$ in the Eq.~\ref{Eq1:magnetization} will be neglected.\cite{komentarz1}
Having in mind the EPR identification of the Fe impurity as the Fe$^{3+}$, with spin $S=5/2$ and $L=0$, we assume $M_{(Zn,Fe)O}(B, T)$ in the form:
\begin{equation}
M_{(Zn,Fe)O}(B, T) = A g \mu_B S B_{S=5/2}(B, T),
\label{Eq2:magnetization}
\end{equation}
where $B_{S=5/2}(B, T)$ is the Brillouin function for spin S~=~5/2, g is g-factor (see Sec.~\ref{sec:EPR}) and A is the number of the Fe centers. Possible interaction between the Fe$^{3+}$ ions is neglected here due to their expected low concentration.
In order to extract $M_{(Zn,Fe)O}(B, T)$ from the total magnetization of the samples $M_{exp}(B, T)$, the dominating diamagnetic contribution (i.e. $\chi_{dia}$ B) is subtracted using the following procedure. At T = 300~K the Brillouin function in Eq.~\ref{Eq2:magnetization} is a linear function of the magnetic field. So that fitting of the M$_{exp}$(B, T = 300~K) by a straight line provides a susceptibility of the sample, which is in practice solely a diamagnetic susceptibility (a residual paramagnetic contribution originating from the Fe$^{3+}$ ions is negligibly small with respect to the total sample susceptibility due to a relatively high temperature and a small density of the Fe$^{3+}$ dopant). The diamagnetic susceptibility evaluated in this way from the data in the field range $2~T < B <7~T$ ($\chi_{dia} = - 4.95 \times 10^{-6}$) is used for evaluation of $M_{(Zn,Fe)O}(B, T)$ according to Eq.~\ref{Eq2:magnetization}.

\begin{figure}
\includegraphics[width=1\linewidth]{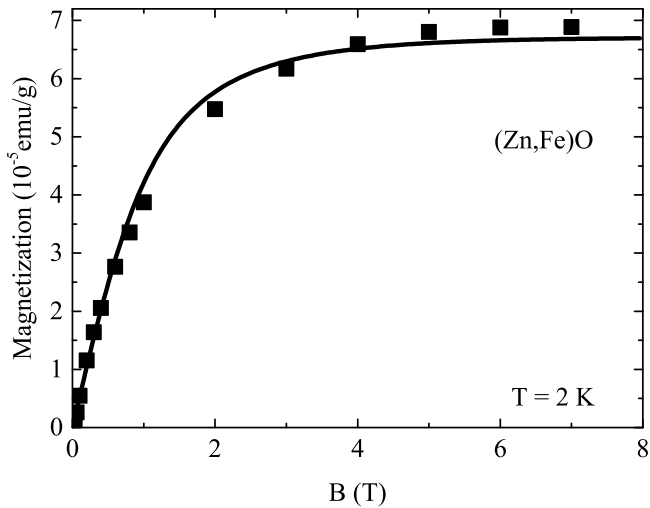}
\caption{The magnetization of the (Zn,Fe)O sample as a function of applied external magnetic field of up to 7~T. Points: experimental data, line: fit following Eq.~\ref{Eq2:magnetization}.}
\label{fig:SQUID}
\end{figure}

In Fig.~\ref{fig:SQUID}, both the experimental data (i.e., $M_{exp}(B,T=2~K)-\chi_{dia}B$) and the fit (i.e., $A g \mu_B S B_{S=5/2}(B, T=2~K)$ are given for an example sample grown on quartz substrate. Parameter $A$ being the only fitting parameter provides the information about the actual molar concentration of the Fe$^{3+}$ ions in the layers. It attains $x = 0.2 \pm 0.05$\%, that is $(5 \pm 1) \cdot 10^{19}$ at/cm$^3$, in agreement with the results of energy-dispersive X-ray (EDX) spectroscopy (not shown). This value is typical for the studied layers. We note that the magnetization measurements confirm the $3+$ state of the iron ions, which points toward high efficiency of the autoionization and/or compensation of the ions predicted by the theory.

\subsection{\label{sec:Transport} DC-resistivity and Hall effect}
Electrical contacts on thin films are deposited using a conductive silver paint. Contact ohmicity is systematically verified by I-V characteristics. DC- resistivity and Hall effect measurements were performed in a Van der Pauw configuration in the temperature range from 90~K to 400~K and for magnetic field perpendicular to the film plane varying from 0 T to 1.6~T using a custom designed, high impedance measurement set-up. This set-up is a combination of Keithley (USA) electrical measurement device and a temperature controller (Linkam Scientific, UK) coupled by a software written in GEMaC, Versailles. The equipment enables measurements of highly resistant samples, up to 10~G$\Omega$.

The samples exhibit a semiconducting behavior. Namely, the resistivity increases with the temperature decrease. At room temperature, resistivity $\rho$ of the 0.2\% (Zn,Fe)O on glass substrate equals to $ 6.6 \times 10^3 \Omega \cdot$~cm. In conductivity \emph{vs} inverse temperature plots, two different regions of the conductivity are distinguished: (i) for 300~K down to 200~K, where conduction due to the temperature activation to the band dominates, (ii) for temperatures below 200~K, where the near neighbor hopping mechanism becomes dominant. The activation energy of band conductivity has been determined to be $0.25 \pm 0.02$~eV.\cite{Chikoidze:JAP2013, Boshta:JMatSci2014} Negative sign of Hall voltage is found for all layers in Hall effect measurements at room temperature. This implies that the samples, independently if deposited on glass or on fused silica, are of n-type. For pure ZnO electron concentration $n = 5.4 \times 10^{15}~cm^{-3}$ and mobility $\mu$ = 1.9~cm$^2$/Vs are found, while for the 0.2\% (Zn,Fe)O $n$ reaches $1.5 \times 10^{16}~cm^{-3}$ and mobility decreases due to the presence of scattering centers down to $\mu$ = 0.5~cm$^2$/Vs. A natural explanation for the increase of the density of free electrons upon the iron doping is that at least some part of iron ions are donors, which are not fully compensated by the native acceptors. In fact, the increase of the electron density from $\sim 10^{15}$~cm$^{-3}$ to $\sim 10^{16}$~cm$^{-3}$ upon the Fe doping is at least two orders of magnitude lower than it would result from the density of the ionized Fe ions ($\sim10^{19}$~cm$^{-3}$ already at T = 2~K, see Sec.~\ref{sec:SQUID}). Thus, we state that most of the ions, which ionize from Fe$^{2+}$ to Fe$^{3+}$ state are compensated by native acceptors. As indicated by previous experimental theoretical~\cite{theo_Lany, theo_Janotti} and experimental~\cite{exp_Tuomisto, Heitz:PRB1992} works, a likely source of native acceptors in ZnO is a Zn vacancy, however a contribution to the carrier concentration may come also from oxygen vacancies\cite{Hulya:PRB2012, Erdem:JofAlloysComp2014, Repp:SpectrchemAPA:2016} or surface states of the studied polycrystalline sample.

\section{\label{sec:Magnetospectroscopy} Experiment: evidence for {\emph{s,p-d}} exchange interaction in {(Zn,Fe)O}}

\begin{figure*}[t!]
\includegraphics[width=0.95\linewidth]{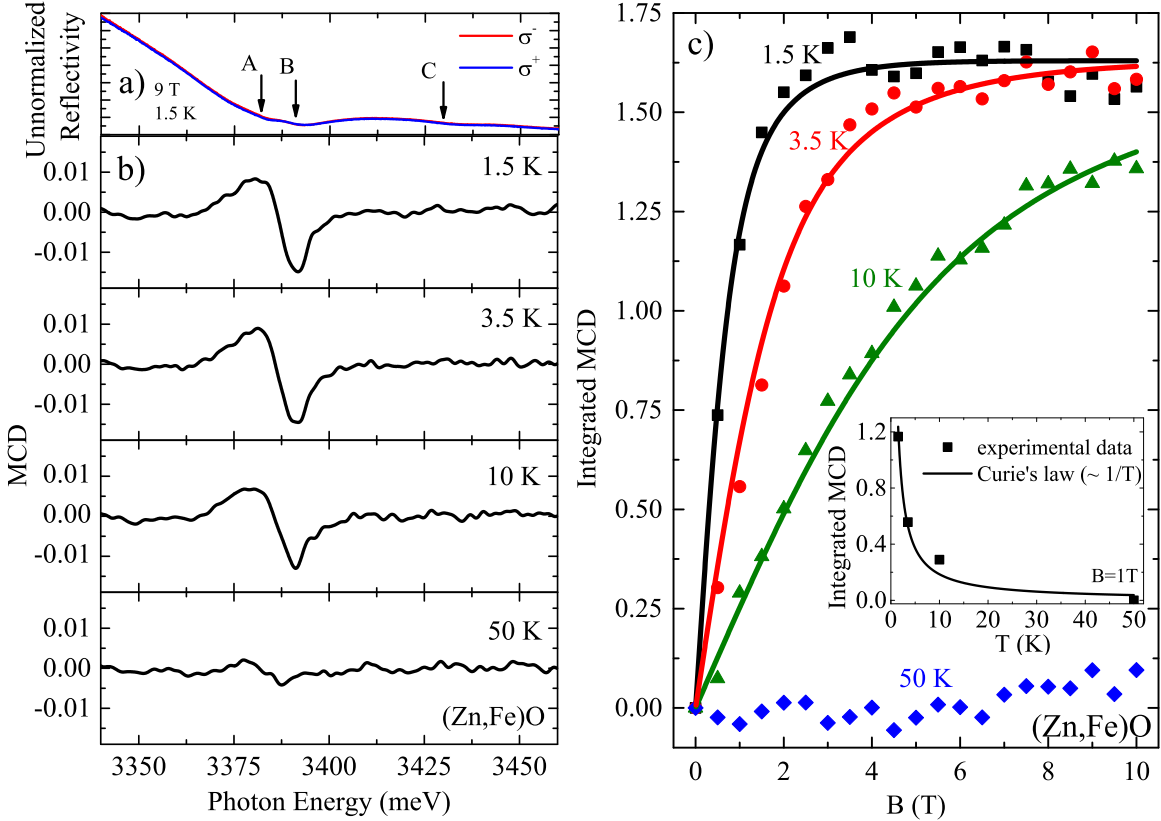}
\caption{(Color on-line) (a) Reflectivity spectra of (Zn,Fe)O at B = 9~T and T = 1.5~K for two circular polarizations of the light. Approximated energies of $A$, $B$ and $C$ excitonic transitions are indicated by arrows. (b) MCD spectra at B = 9~T for consecutive temperatures. (c) Integrated MCD at T = 1.5 K, 3.5~K, 10~K and 50~K plotted as a function of magnetic field of up to 10~T. Inset: Integrated MCD at B = 1~T \textit{vs} the temperature (points) along with the fit by $\sim1/T$ (solid line).}
\label{fig:MCD}
\end{figure*}

The samples are placed at pumped helium temperature inside a cryostat equipped with a superconducting coil magnet. The emission is non-resonantly excited at 3.81~eV (325~nm) using a continuous wave He-Cd laser. The excitation beam is focused to a 0.1~mm spot on the sample surface. The signal is detected by a CCD camera coupled to a grating spectrometer (0.1~meV of overall spectral resolution of the setup). Circular polarizations of the signal are resolved.

The reflectivity and PL measurements are performed in the Faraday configuration in magnetic field of up 10~T, in temperature range from 1.5~K to 50~K, with a halogen lamp serving as a source of the ultraviolet light. The MCD is determined based on the acquired reflectivity spectra as $MCD = (R_{\sigma+}$ - R$_{\sigma-})/(R_{\sigma+}$ + R$_{\sigma-})$, where R$_{\sigma+}$ and R$_{\sigma-}$ represents intensity of the reflectivity spectrum in the $\sigma+$ and $\sigma-$ polarization, respectively.

\subsection{\label{sec:MCD} Magnetic Circular Dichroism}

Optical transitions of three excitons, $A$, $B$ and $C$, are present in the reflectivity spectra of the studied (Zn,Fe)O layers, as expected for a wurtzite structure semiconductor (see Fig.~\ref{fig:MCD}(a). Magnetic field-induced splittings are small with respect to the transitions linewidths, what precludes tracing of the excitonic shifts as a function of the applied magnetic field. However, a clear MCD signal related to excitons is observed in the magnetic field. The MCD spectra determined based on the reflectivity spectra at B = 9~T for temperatures of 1.5~K, 3.5~K, 10~K and 50~K are shown in Fig.~\ref{fig:MCD}b.

Integrated MCD intensity\cite{Rousset:PRB2013} (I$_{MCD}$) is calculated as an integral under the MCD curve in the region of $A$ and $B$ excitonic transitions. The I$_{MCD}$ increases with the magnetic field following a Brillouin-like dependence with a saturation at around 3~T and around 7~T for 1.5 K and 3.5 K, respectively (see Fig.~\ref{fig:MCD}(c)). As it is seen, the I$_{MCD}$ is well described by the paramagnetic Brillouin function with the Land\'{e} factor g = 2.0062 (taken following Ref.~\onlinecite{Acikgoz:JPCM2014}) and spin 5/2 (as for the Fe$^{3+}$ ions), without any free fitting parameters (saturation value of the I$_{MCD}$ is determined once for the 1.5~K case and then kept constant in the case of fits for 3.5~K and 10~K.)
The Brillouin like shape of the observed dependence proves that Zeeman splitting of bands yielding the I$_{MCD}$ results directly from the \emph{s-d} and \emph{p-d} exchange interactions between the iron ions spins (S = 5/2) and spins of, respectively, band electrons and holes. When the magnetic moments of the iron ions become fully oriented at a sufficiently high field, the exchange interaction induced splitting, and thus the I$_{MCD}$ saturates, as observed. In the case of a reference sample of pure ZnO a linear dependence of the I$_{MCD}$ on magnetic field originating form the Zeeman splitting of bands is found, as expected (not shown).

Below the saturation, the MCD magnitude strongly decreases with the increasing temperature following a Curie paramagnetic dependence, 1/T (see inset to Fig.~\ref{fig:MCD}(c) showing the I$_{MCD}$ as a function of the temperature at B = 1~T.) If the exchange interaction of carriers with Fe$^{2+}$ ions played a dominant role in the studied samples, the Van Vleck type paramagnetism, proper for an ion with 3(d$^{6}$) electron configuration, thus with a degenerate ground state, would dominate.\cite{Smolenski:NatureCommun2016, VanVleck:book1932, Mahoney:JChemPhys1970, Mauger:PRB1991, BenoitlaGuillaume:Springer1991} In that case the decrease of the I$_{MCD}$ with the temperature would be only weak in the considered temperatures range. Hence, the observed rapid drop of the I$_{MCD}$ with the increasing temperature indicates that abundance of the Fe$^{2+}$ ions in our samples is negligible and/or that the magnitude of the \emph{s,p-d} interaction is much larger in the case of the Fe$^{3+}$ than of the Fe$^{2+}$ ions. Since the theory (see Sec.~\ref{teo3}) points towards carrier-ion exchange constants comparable or larger for the Fe$^{2+}$ than for Fe$^{3+}$ ions, we state that a content of the Fe$^{2+}$ ions is at least order of magnitude lower than the one of the Fe$^{3+}$ ions. This result is consistent with the magnetometry results at T~=~2~K, which are properly described without taking into account any contribution from the Fe$^{2+}$ ions (Sec.\ref{sec:SQUID}).

Concluding this Section, the MCD provides a strong confirmation for the presence of the Fe$^{3+}$ ions, negligible density of the Fe$^{2+}$ ions as well as for the \emph{s,p-d} interaction between the Fe$^{3+}$ ions and the band carriers in the (Zn,Fe)O.

\subsection{\label{sec:PL} Magneto-Photoluminescence}
\begin{figure*}[t!]
\includegraphics[width=0.95\linewidth]{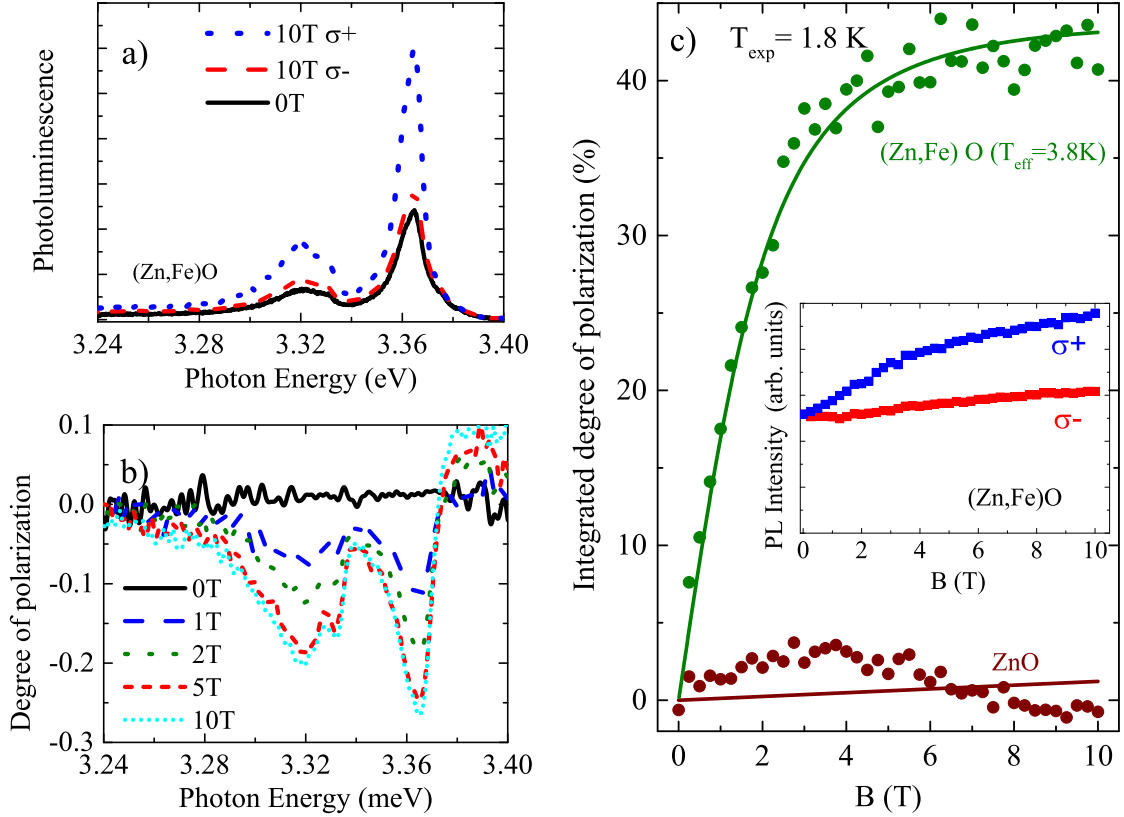}
\caption{(Color on-line) (a) Photoluminescence spectra in magnetic field B~=~0~T and B = 10~T recorded in two circular polarizations $\sigma+$ and $\sigma-$, (b) Degree of Circular Polarization determined from the PL spectra and (c) Integrated Degree of Circular Polarization as a function of magnetic field of up to 10~T for (Zn,Fe)O sample and a reference, a pure ZnO sample (points) along with the fit (lines, see text for details).}
\label{fig:PL1}
\end{figure*}

Figure~\ref{fig:PL1}(a) shows PL spectra in magnetic field B = 0~T and B = 10 T recorded in two circular polarizations of the light, $\sigma+$ and $\sigma-$, for the (Zn,Fe)O sample. The contributions from a bound exciton at around 3360~meV and donor acceptor pairs (DAP) at around 3320~meV 
are identified following Ref.~\onlinecite{Meyer:pssb2004}.

Figure~\ref{fig:PL1}(b) shows a degree of the circular polarization $P$ determined from spectra in magnetic field from 0~T to 10 T for (Zn, Fe)O samples. The $P$ is determined as: $P = (I_{\sigma+}$ - I$_{\sigma-})/(I_{\sigma+}$ + I$_{\sigma-})$, where I$_{\sigma+}$ and I$_{\sigma-}$ represent intensity of the PL spectrum in the $\sigma+$ and $\sigma-$ polarization, respectively. A high degree of polarization $P$ is observed in the spectral region of the bound exciton and of the DAP at energy of around 3365~meV and 3320~meV, respectively. The $P$ increases with the magnetic field and saturates at around B = 5~T.

Integrated degree of polarization I$_P$, defined as the area under the degree of polarization curve in the excitonic region, is plotted as a function of the magnetic field in Fig.~\ref{fig:PL1}(c) together with a paramagnetic Brillouin function fit. The fitting parameters are the saturation value and the effective temperature. The best fit is obtained for factor g = 2.0062,\cite{Acikgoz:JPCM2014} the effective temperature T = 3.8~K and spin 5/2 as for the Fe$^{3+}$ ions. As it is seen the dependence is well described by the Brillouin function. This strongly suggests that the $P$ is directly proportional to the sample magnetization. We have checked that the increased temperature with respect to the temperature of experiment (1.8~K) results from the sample heating with the excitation beam. A dependence on magnetic field is qualitatively the same also for the DAP (not shown) indicating that qualitatively the same mechanism is responsible for the effects observed for the bound exciton and the DAP.

A Lorentzian curve is fitted to the exciton transition at around 3360~meV, providing information on parameters of the transition as a function of the magnetic field. The PL intensity determined that way for the (Zn,Fe)O for $\sigma+$ and $\sigma-$ polarizations is shown in the inset to Fig.~\ref{fig:PL1}(c). The PL intensity increases with the magnetic field in both polarizations. A similar effect on the field was previously observed in DMS structures and nanostructures, e.g., involving Mn.~\cite{Abramishvili:SSC1991, Nawrocki:PRB1995, Lee:PRB2005, Pacuski:PRB2011} As in previous works,~\cite{Abramishvili:SSC1991, Nawrocki:PRB1995, Lee:PRB2005, Pacuski:PRB2011, Galkowski:JAP2015} we attribute the emission intensity increase to the magnetic field induced reduction of efficiency of non-radiative exciton recombination assisted by Auger excitation of the Fe$^{3+}$ ion. In the absence of the magnetic field, band carriers and the Fe$^{3+}$ ions are spin degenerate making efficiency of the Auger process independent of spin. However, the magnetic field lowers the number of spin arrangements of the electron and the Fe$^{3+}$ ion that fulfill the spin conservation rule in the process, and thus it decreases the overall efficiency of the non-radiative Auger recombination. We note that the $\sigma+$ polarized emission increases more than the $\sigma-$ polarized one. Similar as in the study of (Zn,Mn)O (Ref.[\onlinecite{Pacuski:PRB2011}]), this effect can be explained in terms of spin-dependent exciton formation involving free and bound exciton states, as well as $A$ and $B$ excitons relaxation involving change of the exciton symmetry. For pure ZnO, the intensity of the excitonic PL linearly increases for $\sigma+$ and decreases for $\sigma-$ in the magnetic field (not shown) due to polarization of the carriers induced by the Zeeman splitting of bands.

We find also that the linewidth of the bound exciton in (the Zn,Fe)O decreases nonlinearly with the magnetic field in both polarizations with a saturation at around 4~T (not shown). The effect is stronger for the $\sigma+$ than for the $\sigma-$ polarization. We attribute the linewidth narrowing to a reduction of the fluctuations of the magnetization of the Fe$^{3+}$ ions induced by the magnetic field.\cite{Brazis:SSC2002} Mutually opposite shifts in the magnetic field of bound excitons originating from bands of symmetry $\Gamma_7$ and $\Gamma_9$ ($A$ and $B$ excitons) also contribute to the linewidth narrowing.\cite{Pacuski:PRB2011}

To summarize this part, the PL results provide a strong support for the conclusions drawn from the reflectivity measurements, unequivocally confirming presence of the ion-carrier \emph{s,p-d} interaction in the studied system.

\section{Summary and Conclusions}
Theoretical and experimental analysis of the electronic structure of Fe ions in ZnO and of the magnetic properties of ZnO:Fe was conducted. The GGA$+U$ method was employed, with the values of $U$ for $d$(Zn) and $p$(O) orbitals fitted to ZnO experimental band structure. The $U$(Fe) was also considered as a free parameter. For small values of the $U$(Fe), $0<U<3$~eV, the calculations lead to unphysical results, with fractional occupancies of the Fe level degenerate with the conduction band. The calculations performed using the $U$(Fe)$>3$~eV indicate that the Fe is a relatively shallow donor with the $2+$ charge state stable in the ideal ZnO. The $U$(Fe)$=4$~eV allows one to reproduce the observed ionization energy $E_{ion}=0.25 \pm 0.02$~eV. The high energy of the $\varepsilon(+/0)$ transition level facilitates compensation of the Fe$^{2+}$ by unintentional acceptors. Thus, in real imperfect samples both, the Fe$^{2+}$ and the Fe$^{3+}$, should co-exist.

In parallel, EPR, magnetometry, reflectivity, magneto-photoluminescence, and conductivity experiments were conducted on polycrystalline ZnO layers with the Fe content of the order of 0.1 of atomic per cent. The results of the EPR measurements reveal the presence of the substitutional Fe with the 3+ valency in the studied samples. The layers magnetization determined in SQUID measurements is well described with the paramagnetic Brillouin function as for the Fe$^{3+}$ ions, what demonstrates that the dominant Fe charge state is 3+. Consistently with this finding, the magnetic field dependencies in magnetooptical response of the samples determined in the reflectivity and PL measurements are proportional to the measured magnetization. Namely, they are described by the paramagnetic Brillouin function determined as for the Fe$^{3+}$ ions and they obey the paramagnetic Curie law, characteristic of the Fe$^{3+}$ in ZnO, rather than the paramagnetic Van Vleck law expected for the Fe$^{2+}$.

According to the conductivity measurements, the samples are n-type, with the room temperature electron concentration of about 10$^{16}$ cm$^{-3}$. The measured activation energy of conductivity of $0.25 \pm 0.02$~eV is in agreement with the theoretical value of the Fe ionization energy for $U$(Fe)=4~eV.

The low electron concentration indicates a high degree of compensation of the Fe donors, most probably by the zinc vacancies.~\cite{Janotti_PRB76, theo_Lany, Oba2008, Look} The presence of compensating acceptors is also reflected by the DAP recombination line seen in the photoluminescence. Overall, the set of experimental results indicates in a consistent way that the content of the Fe$^{2+}$ ions in the studied samples is at least order of magnitude lower than the one of the Fe$^{3+}$ ions.

Finally, the calculations indicate that the exchange interaction coupling constant $N_0\alpha$ between conduction electrons and the Fe$^{2+}$ or the Fe$^{3+}$ ions, is around 0.4~eV, somewhat higher than values typical for wide band gap semimagnetic semiconductors. The exchange coupling constant $N_0\beta$ for holes strongly depends on the Fe charge state: for the Fe$^{2+}$ the $N_0\beta$ is of the same order as reported for other wide gap DMSs (around 2~eV and ferromagnetic), while in the case of the Fe$^{3+}$ it changes its  sign to antiferromagnetic and is much lower (equal to around -0.4~eV). The pronounced magnetooptical effects observed in the MCD and the PL unequivocally confirm a presence of the \emph{s,p-d} interaction between the band carriers and the Fe$^{3+}$ ions in the studied system. The properties resulting from the Fe doping indicate the (Zn,Fe)O as a promising material for implementation in functional spintronic devices.

\section*{Acknowledgments}
One of the authors (JP) acknowledges the support from Preludium project nr DEC-2013/11/N/ST3/04062, and two authors (AC and PB) the support from the project nr 2012/05/B/ST3/03095, which are financed by Polish National Science Centre (NCN).
Calculations were performed on ICM supercomputers of University of Warsaw (Grant No. G46-13 and G16-11). We thank dr Andrzej Witowski for performing FIR spectroscopy measurements.

\appendix\section{\label{Appendix}}
\begin{figure}[t]
\begin{center}
\includegraphics[width=1\linewidth]{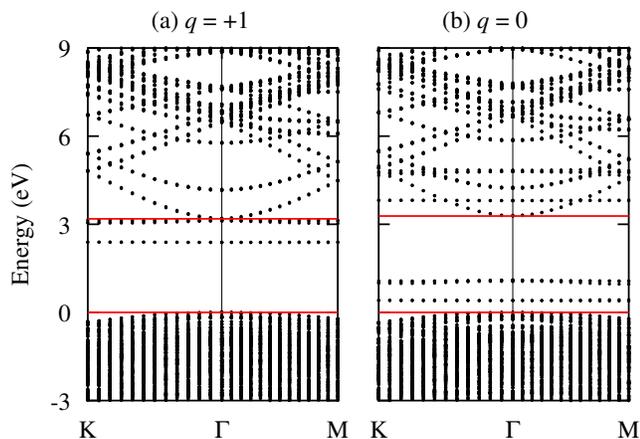}
\caption{\label{figA1}
(Color on-line)
Energy bands of ZnO with Fe in: (a) $q=+1$ charge state {Fe$^{3+}$}, and (b) $q=0$ charge state, "Fe$^{2+}$". Red lines denote the band gap of ZnO. Zero energy is set at the VBM. Results are obtained with the 144-atom supercell for $U$(Fe)=0.}
\end{center}
\end{figure}
Here we analyze electronic structure of Fe in ZnO calculated assuming small $U$(Fe), and show problems encountered within GGA. The band structure and DOS of ZnO doped with Fe$^{3+}$ and Fe$^{2+}$ is presented in Fig.~\ref{figA1}(a) and Fig.~\ref{figA1}(b), respectively, for $U$(Fe)$=0$. The impurity $d$(Fe) states constitute nearly flat bands, and their energies strongly depend on the Fe charge state. Fe$^{3+}$ introduces two spin-down states close to the CBM: the $e_{2\downarrow}$ doublet is at 2.4~eV, and the $t_{2\downarrow}$ triplet is very close to the CBM. Both spin down levels are empty.

Addition of one electron to the ZnO:Fe$^{3+}$ system should result in the $q=0$ charge state Fe$^{2+}$, with 6 electrons on the $d$(Fe)-induced levels. However, this configuration cannot be reached.  The electronic structure for $q=0$ calculated allowing for fractional occupancies is shown in Fig.~\ref{figA1}(b). In this case, the Fe spin-down states are degenerate with the conduction band. In particular, the $e_{2\downarrow}$ doublet is about 0.5~eV above CBM, and $t_{2\downarrow}$ is higher by $\sim 1$~eV. Therefore, the electron that should occupy $e_{2\downarrow}$ autoionizes to the bottom of the conduction band. A more detailed analysis shows that autoionization is partial only.
In fact, the calculated occupancy of $e_{2\downarrow}$ is 0.7, and the integrated occupancy of the conduction states up to the Fermi energy $E_F$ is 0.3. One should notice that since there are no empty states below $E_F$, $E_F$ must be located within the partially occupied $e_{2\downarrow}$ level. This electronic configuration for $q=0$, with partial occupation of $e_{2\downarrow}$, denoted as "Fe$^{2+}$", is different from both the Fe$^{3+}$ state, and from Fe$^{2+}$ with one electron on $e_{2\downarrow}$.

One should observe, however, that the results for "Fe$^{2+}$" are largely artefacts which follow from the employed method of calculations. More specifically, there are two issues: the increase of eigenenergies with increasing occupation, and applicability of partial occupations approach. First, the calculated increase of the Kohn-Sham energy of $e_{2\downarrow}$ with its increasing occupation is related to the spurious self-interaction obtained in the approximate versions of DFT. Within the exact DFT, a Kohn-Sham energy level is independent of its fractional occupation, while an integer change of occupation induces its jump. Indeed, a change of the charge state of a defect due to a capture (emission) of an electron implies an increased (decreased) Coulomb repulsion between electrons situated at the relevant orbitals ($d$ orbitals of a transition metal ion, broken bonds of vacancy neighbors, etc.).
This is reflected in the increasing energies of transition levels for consecutive charge states, together with the increasing Kohn-Sham eigenenergies of defect-induced states. This is a genuine physical effect, which can be described by the so-called effective $U$ parameter, and calculated within DFT~\cite{VdW} for integer occupancies. Second, the final result for "Fe$^{2+}$" with one electron shared between $d$(Fe) and CBM should not be accepted because such a partition is not physical in the present case of an isolated defect center in an insulator.

In other words, the $d^6$ configuration of Fe$^{2+}$ is not stable since $e_{2\downarrow}$ occupied with one electron is above the (empty) CBM,~\cite{Footnote} similar to a resonant donor. However, a complete autoionization to the $q=+1$ charge state cannot occur because the empty $e_{2\downarrow}$ is below the (occupied) CBM. Neither case corresponds to the ground state, and the convergent self-consistent solution is only obtained for fractional occupation of the $e_{2\downarrow}$ level. These problems with "Fe$^{2+}$" in ZnO are analogous to the difficulties encountered for isolated transition metal atoms such as Fe.~\cite{Janak, Koch} Namely, the self-consistent solution for $d^ns^1$ configuration leads to an empty $s$ state below the partially filled $d$ states, while in the $d^{n-1}s^2$ configuration there are empty $d$ states below the $s$ level.
The energy minimum is obtained for fractional occupations of both $d$ and $s$, but, as indicated by Janak,~\cite{Janak} "In a strict interpretation of the Kohn-Sham theory, this means that there is no charge density for this problem which minimizes the total energy and which is generated from a Fermi distribution."

The electronic structure of the neutral Fe in ZnO calculated in Refs~\onlinecite{sandra, spaldin, t11} is characterized by $E_F$ in the conduction band, and partially filled $e_{2\downarrow}$, which closely resembles the "Fe$^{2+}$" configuration. Clearly, without a detailed inspection of the results it is not possible to assess the genuine resonant (or the "pseudo-resonant") character of Fe found in those works but, given the similarity to "Fe$^{2+}$", one may expect that the above discussion may be relevant in some of those cases.

\bibliographystyle{apsrev_my}

\begin{thebibliography}{114}
\bibitem[{ref()}]{ref_occupancy}
\bibinfo{note}{In the case of large values of $U$(Fe), the equilibrium
  electronic configurations are sensitive to the initial occupancies of $d$(Fe)
  orbitals, and all possible cases should be considered to obtain the correct
  solution. The lowest total energy of Fe$^{2+}$ is obtained for the fully
  occupied $a_{t\downarrow}$.}
\bibitem[{bon()}]{bonds}
\bibinfo{note}{The localized character of $d$(Fe) is responsible for the
  relatively large changes of bond lengths with both the charge state and
  $U$(Fe). Because of the wurtzite symmetry, the Fe-O bond $d_1$ along the
  $c$-axis is not equivalent to the three remaining basal bonds $d_2$. The
  calculated values for $U$(Fe)=0 are $d_1=1.89$ (1.96)~\AA\ and $d_2=1.89$
  (1.95)~\AA\ for Fe$^{3+}$ ("Fe$^{2+}$"), respectively. The increase of bond
  lengths by about 3.5~\% induces an upward shift of $e_{2\downarrow}$ and
  $t_{2\downarrow}$ thanks to the antibonding character of their wave
  functions. Using $U$(Fe) of 4-5~eV makes the $d$ orbitals more localized,
  which in turn affects the bond lengths. In this case, the bonds increase by
  about 6~\% to $d_1=1.91$ (2.15)~\AA\ and $d_2=1.90$ (2.00)~\AA\ for Fe$^{3+}$
  (Fe$^{2+}$).}
\bibitem[{not()}]{note4}
\bibinfo{note}{We checked that changes by $\pm 1$~eV of $U$(Zn) and $U$(O)
  fitted in Sec.~\ref{teo1} affect the fitted value of $U$(Fe)=4 eV by about
  0.3 eV, but the overall picture, including problems with "Fe$^{2+}$",
  remain.}
\bibitem[{kom()}]{komentarz1}
\bibinfo{note}{We note that even in the case C is nonzero, its role in
  Eq.~\ref{Eq1:magnetization} is minimized by a procedure we use to extract
  $M_{ZnO:Fe}(B, T)$.}
\bibitem[{Foo()}]{Footnote}
\bibinfo{note}{The Kohn-Sham energy of $e_{2\downarrow}$ occupied with one electron
  (while keeping the conduction band empty) was not calculated directly, since
  the calculations do not converge. Its approximate energy is 1 eV above CBM,
  but the exact value is not needed to assess the "pseudo-resonant" character
  of Fe in ZnO.}
\end{thebibliography}

\end{document}